# Quantum-Secure Authentication with a Classical Key

*Sebastianus A. Goorden,[1], Marcel Horstmann[1,2], Allard P. Mosk[1], Boris Škorić,[3], and Pepijn W.H. Pinkse[1,4]*

**Authentication provides the trust people need to engage in transactions. The advent of physical keys that are impossible to copy promises to revolutionize this field. Up to now, such keys have been verified by classical challenge-response protocols [1, 2, 3, 4, 5, 6, 7, 8]. These protocols are in general susceptible to digital emulation attacks, in which the adversary intercepts the challenge and sends the correct response without needing access to the physical key. Here we demonstrate Quantum-Secure Authentication ("QSA") of an unclonable physical key (a classical multiple-scattering medium) using weak coherent light pulses. The authentication process is inherently secure by virtue of quantum-physical principles [9, 10, 11, 12]. QSA operates in the limit of a large number of channels, represented by the more than thousand degrees of freedom of an optical wavefront shaped by a spatial light modulator [13]. In contrast, the light pulse probing the key contains only few photons. Since the shape is unpredictable for an adversary, the availability of very few photons compared to the number of channels makes it impossible for him to determine the shape; this provides unconditional security against digital emulation.**

Authentication is an important cornerstone of security. Authentication of persons can be based on "something that you know", e.g. digital keys, or "something that you have", e.g. physical objects such as classical keys or official documents. A drawback of digital keys is that their theft can go unnoticed, but also the copying of a traditional physical key can be done secretly. A Physical Unclonable Function (PUF) is a physical object that cannot feasibly be copied because its manufacture inherently contains a large number of uncontrollable degrees of freedom [1, 3]. A PUF is a function in the sense that it reacts to a stimulus ("challenge") by giving a response. After manufacture there is a one-time characterization of the PUF in which its challenge-response behavior is stored in a database. The PUF (from this point referred to as the "key") can later be authenticated by comparing its response behavior to the database, see Fig. 1a.

[1] Complex Photonic Systems (COPS), MESA+ Institute for Nanotechnology, University of Twente, PO Box 217, 7500 AE Enschede, The Netherlands
[2] Laser Physics and Nonlinear Optics, MESA+ Institute for Nanotechnology, University of Twente, PO Box 217, 7500 AE Enschede, The Netherlands
[3] Eindhoven University of Technology, PO Box 513, 5600 MB Eindhoven, The Netherlands
[4] Applied Nanophotonics, MESA+ Institute for Nanotechnology, University of Twente, PO Box 217, 7500 AE Enschede, The Netherlands.

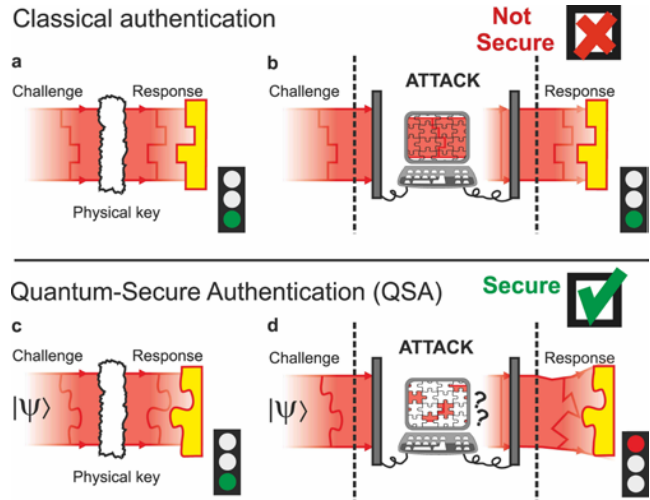

Figure 1| The idea of Quantum-Secure Authentication (QSA): **a**, In classical authentication of an optical unclonable physical key, a challenge wavefront of sufficient complexity is sent to the key. The response wavefront is compared with those stored in a database (yellow pieces) to make a pass (green light) or fail (red light) decision. However, this verification can be spoofed by an emulation attack (**b**) in which the challenge wavefront is completely determined and the expected response is constructed by the adversary who knows the challenge-response behavior of the key. In Quantum-Secure Authentication (**c**) the challenge is a quantum state for which an emulation attack (**d**) fails because the adversary cannot actually determine the quantum state and hence any attempt to generate the correct response wavefront fails.

PUFs are used to prevent *physical emulation* of the key, i.e., making a sufficiently accurate clone or concocting a device that mimics its *physical* behavior. This is infeasible, though not theoretically impossible, given the properties of PUFs [1, 3]. See also the supplementary material. However, PUFs are in general still vulnerable to a class of attacks that we will refer to as *digital emulation* (Fig. 1b). Here the adversary has knowledge of the key's properties either from physical inspection of the key or by access to the challenge-response database. He intercepts challenges and is able to provide the correct responses by looking them up in his database. This is a highly relevant scenario as accessible databases are notoriously difficult to protect. So far the only defense against digital emulation is to deploy various sensors that try to detect if some form of spoofing is going on. This leads to an expensive arms race in which it is difficult to ascertain the level of security.

In this paper we present Quantum-Secure Authentication (QSA) of optical keys, a scheme with highly desirable properties: QSA
- uses a key that is infeasible to emulate physically.
- is unconditionally secure against digital emulation attacks.

- does not depend on secrecy of any stored data.
- does not depend on unproven mathematical assumptions.
- is straightforward to implement with current technology.

The use of quantum physics in QSA is inspired by quantum cryptography [9, 10, 14]. However, there are major differences. The aim of quantum cryptography is to generate a secret digital key known only to Alice and Bob, whereas QSA allows Alice to check if Bob possesses a unique physical object. Quantum cryptography requires the existence of an authenticated channel between Alice and Bob, typically based on a secret key that is shared beforehand [15]. In contrast, QSA needs only publicly available information; there are no secrets. See the supplementary material for an overview of cryptographic primitives and their properties.

Our implementation of QSA uses random scattering media as PUF [1, 16, 17]. The challenges are high-spatial-dimension states of light [18, 19, 20] with only a few photons. The response is speckle-like and depends strongly on the challenge and the positions of the scatterers. Due to the noncloning theorem [21] it is impossible for an adversary to fully determine the challenge and therefore to construct the expected response (Fig. 1c-d). The verifier can, however, easily verify the presence of the encoded information with an appropriate basis transformation, authenticating the key.

After its manufacture, the key is enrolled: the challenge-response pairs are measured with as much light as needed. Each of our challenges is described by a 50×50 binary matrix. Each element corresponds to a phase of either 0 or π. A spatial light modulator (SLM1) is used to transform the incoming plane wavefront into the desired challenge wavefront. The challenge is sent to the key and the reflected field is recorded in a phase-sensitive way. The challenge along with the corresponding response is stored in a challenge-response database. In our current implementation this requires 20 kB of computer memory per challenge-response pair which corresponds to 50 MB for a fully characterized key.

After enrolment, keys are authenticated using the setup illustrated in Fig. 2. Our light source is an attenuated laser beam chopped into 500 ns light pulses each containing $n = 230 \pm 40$ photons. Quantum readout of optical keys can be achieved with single or bi-photon states [22], squeezed states [23] or other fragile quantum states [11]. We use coherent states of light with low mean photon number [24], because in QSA they provide a similar security as other quantum states and are easier to implement in real-life applications. A challenge-response pair is constructed using information from the database. SLM1 is used to shape the few-photon challenge wavefront, which is then sent to the key. The reflected wavefront is sent to SLM2, which adds to it the conjugate phase pattern of the expected response wavefront. Therefore, SLM2 transforms the reflected speckle field into a plane wave only when the response is correct. In case the response is wrong, SLM2 transforms the field into a completely different speckle field. When the response is correct, the lens positioned behind SLM2 focuses the plane wave to a point in the analyser plane, as shown in Fig. 2b. A false key will result in a speckle on the analyser plane as shown in Fig. 2c. Compared to the typical peak height in Fig. 2b of 1000 times the background, the loss of intensity in the center of Fig. 2c is dramatic. We spatially filter the field in the analyser plane with a pinhole and image it onto a photon-counting detector. In Fig. 3a we show the typical photodetector signal for the correct response and for an incorrect response provided by the true and a false key, respectively. Only with the true key multiple photodetections are seen. After repeating the measurement 2000 times, Fig. 3b shows the histogram of the number of photodetections for the true key, resembling a Poissonian distribution with a mean of 4.3. Fig. 3b also shows the average histogram of photodetections when 5000 random challenges are sent to the key, with the key and SLM2 kept unchanged. This experiment gives an upper bound on the photodetections in case of an attack with a random key. This histogram resembles a Poissonian distribution with a mean of 0.016 photodetections. We can clearly discriminate between true and false keys.

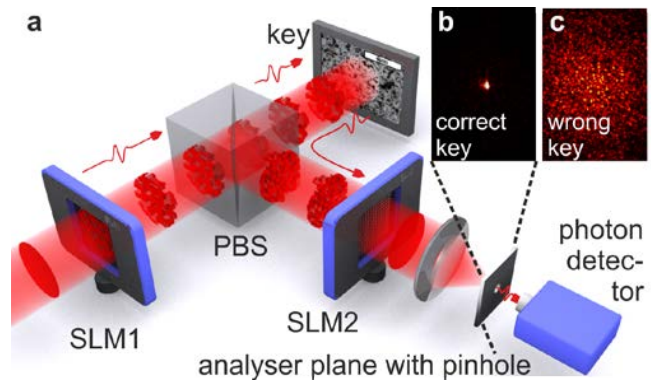

Figure 2| Quantum-secure optical readout of a physical key. **a**, Setup: A spatial light modulator (SLM1) creates the challenge by phase shaping a few-photon wavefront. In the experiment a 50x50 binary phase pattern is used with 0 and π phase delays. The challenge is sent to the ZnO key (scale bar is 4 μm) by a microscope objective (not shown). The response is coupled out by a polarising beam splitter (PBS). The response is transformed back by SLM2 and then focused onto the analyser plane. **b**, Only if the key is the true unique key, the response has a bright spot in the centre, holding ≈60% of the power in the image and allowing that fraction to pass a pinhole and land on a detector where photodetection clicks authenticate the key. **c,** In case of a false key, the response in the analyser plane is a random speckle pattern.

In order to characterize the achievable security for one repetition of our readout, we introduce the quantum security parameter S,

$$S \equiv K/n, \qquad (1)$$

as the ratio of the number of controlled modes $K$ and the average number of photons $n$ in the challenge. The parameter $K$ quantifies the dimensionality of the challenge space and is equal to the number of independent response wavefronts that are obtained by sending in different challenge wavefronts. It is well approximated by the number of speckles on the key illuminated by the challenge [25]. In our experiment we have $K = 1100\pm200$ and $n = 230\pm40$, yielding $S = 5\pm1$. Because a measurement of a photon can extract only a limited amount of information, a large $S$ implies that the adversary can only obtain a small fraction of the information required to characterize the challenge. Therefore he cannot determine the correct response. An adversary who measures an optimal choice of field quadratures of the challenge cannot achieve a fidelity better than approximately [26]

$$F = F_{OK}/(S+1), \qquad (2)$$

where $F$ is the fraction of photons detected by the verifier's hardware in case of an attack and $F_{OK}$ is the fraction of photons detected when the response is correct. This result holds for $S > 1$ and $K \gg 1$ and is in line with the intuition that a measurement of $n$ photons can only provide information about $n$ modes. Operating the readout in the regime $S > 1$ therefore gives the verifier an eminent security advantage which has its origin in the quantum character of light.

In the verification we aim to discriminate a correct key from an optimal attack. Given a conservative lower bound of $S = 4$, the number of photodetections on the single-photon detector in a single readout in case of an optimal (digital emulation) attack follows a Poissonian distribution with mean 0.86, as shown in Fig. 3b. Choosing a threshold of 3 or more photodetections for accepting the key, we find that the measured false reject ratio is 9%. In case of random challenges the false accept ratio is $1.7\times10^{-4}$ % and the theoretical maximum false accept probability in case of the digital emulation attack (Eq. 2) is 6% (Fig. 3c). The security improves exponentially by repeating the verification, every time choosing a different challenge and its corresponding SLM2 setting from the database. The individual photon counts are added, and a combined threshold is set. As illustrated in Fig. 3d, after 10 repetitions the false accept and false reject probabilities are of order $10^{-4}$. As detailed in the supplementary information, after 20 repetitions they are both of order $10^{-9}$. Thus, the false decision rates can be made negligible in a small number of repetitions.

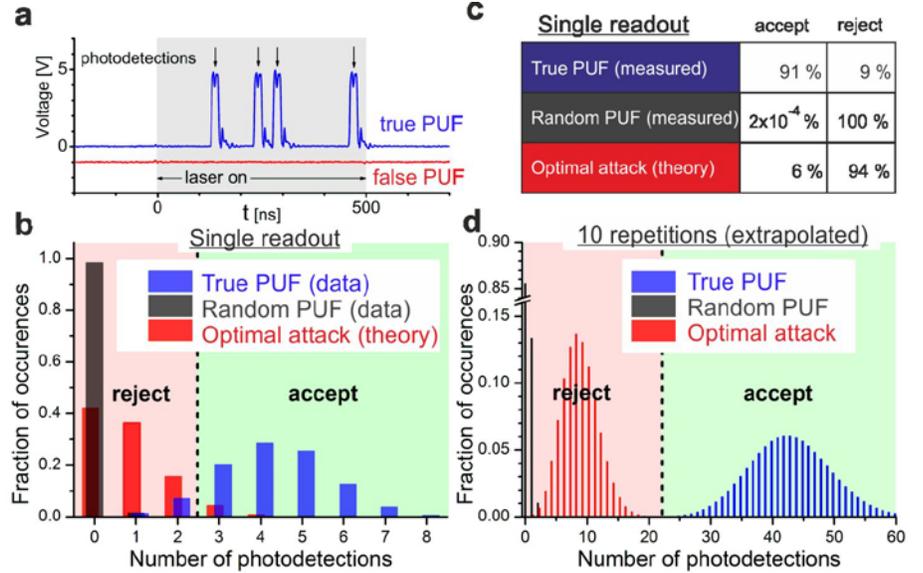

Figure 3| Quantum-secure readout of an unclonable physical key (PUF), using challenge pulses with 230±40 photons distributed over 1100±200 modes. **a**, Real-time examples for the true key (blue line) and a false key (red line, offset for clarity). **b**, Measured number of photodetections in case of the true key, a random key (imitated by sending random challenges to the same key), and for an optimal attack given $S=4$. The threshold is chosen such that the false positive and negative probabilities are approximately equally small assuming an optimal attack. **c**, Acceptance and rejection probabilities in case of the true key, a random key and in case of an optimal digital emulation attack. **d**, Number of photodetections extrapolated to 10 repetitions: the false positive and false negative probabilities quickly decrease to order 0.01 %.

In our implementation, the time for readout is limited to about 100 ms by the switching time of the SLM. Using faster micromirror-based SLMs [27, 28], the complete authentication protocol with 20 repetitions can be performed in less than a millisecond. The one-time enrolment of the key then takes on the order of a second. Quantum-secure authentication does not require any secret information and is therefore invulnerable to adversaries characterizing the properties of the key ("skimming"). Hence, QSA provides a practical way of realizing unprecedentedly secure authentication of IDs, credit cards, biometrics [29] and communication partners in quantum cryptography.

**Appendix**

**Shaping the challenge and response wavefronts**

Two halves (referred to as SLM1 and SLM2) of the same reflective Holoeye HEO 1080P phase-only spatial light modulator are used to shape challenge wavefronts and decode response wavefronts, respectively. We use 50x50 segments, each consisting of 16x16 pixels, to shape challenge wavefronts. Since the used beams are cylindrical, the corners of the 50x50 segments area are not illuminated, so that we effectively use $2500\pi/4 = 1963$ segments of SLM1. Each segment is set to a phase of either 0 or $\pi$, allowing a total of $2^{1963} = 10^{591}$ different challenges, of which 1963 are orthogonal. This number is slightly larger than the number of modes that is supported by the sample area that we illuminate, which is experimentally verified to be 1100±200. The challenge is focused onto the PUF using a 0.95 NA 63x Zeiss microscope objective. The response is collected using the same objective and measured in 130x130 segments using standard phase-shifting interferometry [30]. This data is used to find the phases of the response wavefront at the 130x130 corresponding segments with in total 800x800 pixels on SLM2. SLM2 is then set to flatten the phase of the response wavefront by adding the conjugate phase to the response wavefront.

**Cryptographic context**

In Table T1 an overview of important cryptographic methods is provided. Note that the overview is not limited to authentication methods.

| Method | Purpose | Technical Requirements | Security Assumptions | | Status |
|---|---|---|---|---|---|
| | | | Physical | Mathematical | |
| BB84 [10] | Key exchange | Authenticated classical communication channel; single photons | none | none | Experimentally demonstrated, commercially available |
| SARG04 [31] | Key exchange | Authenticated classical communication channel; weak coherent pulses | none | none | Commercially available, more practical than BB84 |
| Diffie-Hellman [32] | Key exchange | Authenticated classical communication channel | Quantum computers are infeasible to build | Discrete Logarithms are difficult to compute | Widely used |
| RSA [33] | Public-key crypto. Encryption; signatures | | Quantum computers are infeasible to build | "RSA Assumption" ~Factoring a product of two large primes is difficult | Widely used |
| McEliece [34] | Public-key crypto. Encryption; signatures | | none | Decoding an unknown linear code is difficult | Well studied; currently considered less practical than other asymmetric crypto |
| Message Authentication Codes (MACs) [35] | Message authentication | Short secret pre-shared key | none | none | Widely used |
| Mechanical key | Authentication; access control | Physical lock & key mechanism needed; key must be distributed first | Inferring key from lock is difficult | none | Ancient, still widely used |
| Quantum money [36, 37] | Anti-counterfeiting | Long-term storage and high fidelity initialization and readout of quantum memory | none | none | Proposed; Very difficult and expensive with current technology |
| Classical PUF authentication [1] | Object authentication, hands-on | Antispoofing detectors against emulation | PUF is physically unclonable | none | Experimentally demonstrated |
| QSA | Object authentication, hands-off | Fewer photons than modes | Lossless implementation of high-dimensional arbitrary unitaries is infeasible | none | New; Proven unconditionally secure against digital emulation [38] |

Table T1. Overview of important cryptographic methods.

Unconditionally secure authentication methods have been proposed that use qubits as a key, e.g. quantum money and a number of methods based on quantum-entangled keys. These methods require quantum memory with long lifetimes that have not been achieved and are infeasible using current technology. Unconditionally secure classical authentication procedures, such as MACs (which are typically used in conjunction with QKD), are usually symmetric and require the distribution and secret storage of classical keys between every pair of parties that must be able to authenticate each other. Asymmetric classical protocols, such as RSA, are much more practical and are therefore widely used for establishing authentication, despite the fact that they are based on unproven mathematical assumptions and therefore not unconditionally secure. Besides that, also asymmetric protocols typically require secret storage of private keys. Classical authentication which uses a Physical Unclonable Function (PUF) as a key has the advantages that it is practically impossible to copy the key and that it is not based on mathematical assumptions. However, every verifier must possess and be able to store secret information about the physical key. If the information about the key leaks, the key can be emulated using digital devices rendering the authentication protocol insecure. This problem is solved by QSA.

**Security analysis**

An adversary who does not have the PUF may attempt several attack strategies. We will address them here and show why they fail.

1) **Blinding attacks**. An additional detector which measures the total intensity outside the pinhole is sufficient to prevent false positive detections in case an adversary floods the system with light. In addition, flooding can be detected by including fake challenges, for which no photon detections are expected. The time needed for one repetition of the procedure is in practice limited only by the switching time of the SLM, on the order of 100 ms for the present SLM. Therefore there is ample room to randomly include fake challenges where (unknown to the adversary) no signal is expected. This also provides security against attacks that trigger the photodetector by non-optical means such as a beam of ionizing radiation. Note that the same requirements hold for QKD. A QKD implementation has successfully been attacked by blinding the detectors [39].

2) A "**Digital Emulation Attack**", in which the adversary attempts to measure the challenge and then estimates the response. As shown in [14] for single photon states and in [26] for quadrature measurements, this is doomed to fail. A newer result [38] shows that when a challenge consists of $n < K$ quanta in the same state, our scheme is secure against *all* challenge-estimation attacks. A quantitative example shows what the adversary can hope to achieve:

In our experiment the lower bound for the quantum security parameter $S$ is 4. Assuming that the adversary has a perfect photon-counting or quadrature measuring camera, the expected squared inner product between the adversary's best estimate and the correct challenge is equal to $1/(S+1) = 1/5$ [26]. He can therefore expect to obtain a number of photon clicks at the detector that is 1/5 times the expected number of clicks with the correct challenge. The expected number of photons for the correct challenge is 4.3, so the attacker will obtain on average 0.8 photon clicks, well below the acceptance threshold of 3.

Experimentally we tested the scaling at the basis of this argument. We parameterize the challenge wavefronts by a $K$-dimensional complex vector $C_0$. This $C_0$ is chosen by the verifier to yield the maximum light power in the focus behind SLM2 given the presence of the true PUF and the setting of SLM2. Now assume $C_0$ is replaced by another wavefront $C_1$. We quantify the proximity of the challenge $C_1$ to the original challenge $C_0$ by the inner product $C_0 \cdot C_1 = \sum_{i=1..K} C_{0,i}^* \cdot C_{1,i}$, where the sum runs over the mode index. The fraction $F$ of the light energy in the focus was experimentally found to scale as $F = 0.6 \, | C_0 \cdot C_1 |^2$, confirming the scaling explained in [26].

3) Making an **exact physical copy** of the PUF. For our key this requires positioning millions of particles with the exact same high refractive index as zinc-oxide and with exactly the correct shapes at exactly the right positions on a nanometer scale. This is not possible with current technology and also not possible in the foreseeable future. To our knowledge no one even tried this.

4) Making a **passive optical device** that emulates the PUF's physical challenge-response behavior.

Since the PUF only realizes a complex linear transformation, one would be tempted to think that it is straightforward to make a passive optical device which does the same optical transformation as the PUF. It is not. The crucial point is that the adversary cannot know which challenge to expect, and therefore can only succeed if his passive optical device produces the correct response *for a large fraction of the challenge space*. In other words, he will have to emulate a large fraction of the optical properties of the PUF into his optical device. This comes close to making a copy of the PUF. A *three-dimensional* random scattering medium with front surface area $A$ contains much more random information than can be encoded in a random scattering surface of the same area $A$. For our sample parameters, a single diffraction-limited spot focused on the surface of the PUF gives rise to a speckle pattern with a Gaussian envelope with a Full Width at Half Maximum (FWHM) of approximately 5μm, containing about $10^2$ speckles. When we illuminate the PUF with a random challenge, the illumination spot is much larger than a diffraction-limited spot. The PUF is now seen to reflect a speckle pattern with a FWHM of about 15μm, containing the equivalent of about $10^3$ speckles. The reflection matrix describing the PUF is nonlocal (i.e., non-diagonal in any representation) as it connects surface points that are spatially separated by up to 5 μm. It is therefore impossible to emulate the PUF with a single scattering surface (e.g., that of an SLM), which would have a local reflection matrix.

An intriguing form of attack would be to make a PUF-emulating hologram into which a large portion of the PUF's reflection matrix is written. Because of the low index of refraction contrast of photorefractive materials, on the order of 0.02 to 0.1 [40], such a hologram must be significantly larger than the true PUF to obtain sufficient reflectivity. Therefore, this form of attack can be easily foiled using a light source with a coherence length of the order of 30 μm, on the order of the average path length photons travel in the PUF. The average optical path in the hologram is much longer than the coherence length so that no speckle pattern will form.

Another intriguing form of attack would be by means of a PUF-emulating nanophotonic network. Since in principle every passive linear optical network can be emulated by a sufficiently complex network of, e.g., beam splitters [41], this is in theory possible. Work by, e.g., Miller *et al.* [42, 43] shows the concepts needed to make such a network. However, an adversary who wants to emulate the PUF functionality needs to program a passive optical device with $K$ modes and $K^2$ connecting elements while keeping all the involved path lengths equal to within the coherence length.

Despite the huge efforts already spent in making linear optical networks for linear optical quantum computing and quantum simulation, state of the art networks have at maximum on the order of 10 connected beam splitters [44] and losses of 0.2 dB per element for waveguide-based beam splitters [45]. Recently larger networks have been built [46], but they only work because of the high tolerance against phase errors possible in the functionality of the realized phased arrays, which does not apply for a PUF emulator. Alternative photonic-crystal-based networks could be smaller [47], but the corresponding losses are even higher. Hence, the extension to $10^6$ connected nodes with overall losses of less than 10dB, extreme phase sensitivity and simultaneously keeping the differential path lengths to within 30 micrometer is still far out of reach of technology. If possible at all in future, it would require $10^6$ optical elements each with a loss of less than 0.01 dB and a tremendous effort to emulate a single PUF.

A note about the scaling: For an $K$-mode PUF, the adversary needs to make a passive optical device with $K^2$ connecting elements. In our demonstration we have used on the order of $10^3$ modes, but increasing this to $10^5$ seems entirely feasible with present technology, given the availability of megapixel SLMs. With the state-of-the-art beam splitters [45], the required network of in the order of $10^{10}$ optical elements would have a loss of 20000dB and cover an area of the order of $1m^2$.

5) A **quantum computer** can emulate our key only if it can perform arbitrary unitary operations on $K$-dimensional quantum states of light with low losses. The technologically most feasible way to do this seems to be through a tunable low-loss $K$-dimensional optical device. Such a device is at least as difficult to build as a passive low-loss $K$-dimensional optical device. This is infeasible for the same arguments presented in point 4) [48].

6) **Hybrid strategies** that mix (complex) passive elements and measuring devices. In the proof of the security against challenge-estimation attacks [38] no assumptions are made about the basis in which the adversary measures the challenge. Even if the adversary can program an arbitrary linear transformation before the measurement, he cannot breach the QSA scheme by estimating the challenge. In fact it is always better for the adversary to use his linear transformation capabilities directly on the challenge state as in attack 4.

**Repetition for exponential security gain**

Fig. S1 shows the calculated probability of false-positive and false-negative decisions as a function of $S$ and the number of repetitions, with the number of modes $K$ kept constant. For each point in Fig. S1 the threshold was chosen in the minimum between the photon detection distributions obtained with the true PUF and the one calculated for the optimal challenge estimation attack. This leads to false positive and false negative probabilities that are approximately equally small. At a moderate $S$ the probability of an erroneous decision is already of order $10^{-4}$ after 10 repetitions. At high $S$ it takes more repetitions to rule out incorrect decisions since high $S$ (at fixed $K$) implies a low photon number $n$. Since the threshold can only be taken at an integer number of photons, one may notice some quantization steps. For larger numbers of repetitions the probability of an incorrect decision is reduced exponentially and can hence be made arbitrarily small.

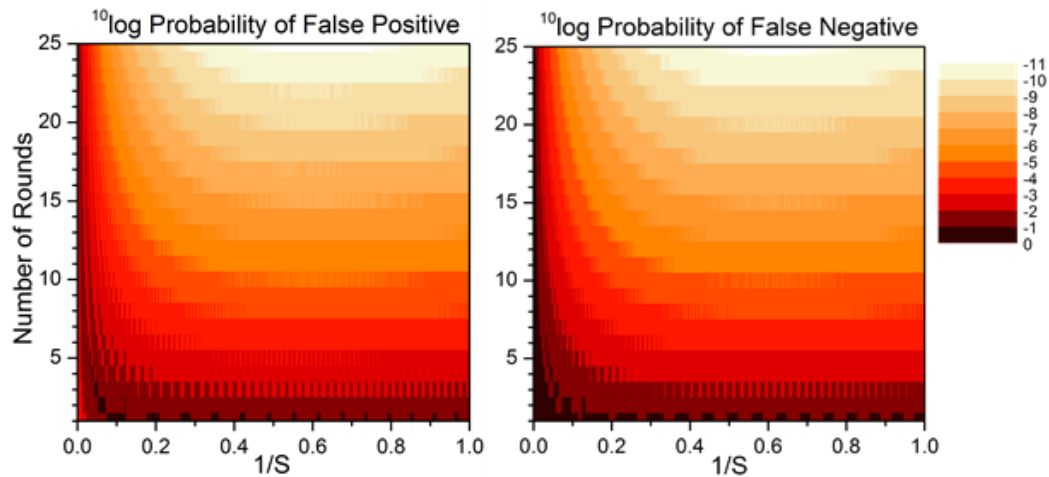

Figure S1. Probability of a false positive (acceptance of a challenge estimation attack) and a false negative decision (rejection of a correct PUF) as a function of the security parameter $S$ and the number of repetitions (rounds). The plot is made by varying $n$ and choosing the optimal threshold, while keeping $K = 1062$.

Correspondence and requests for materials should be addressed to P.W.H.P. (P.W.H.Pinkse@utwente.nl).

**Acknowledgements** We thank J. Bertolotti, K.-J. Boller, G. Giedke, J. Herek, S.R. Huisman, T.J. Huisman, B. Jacobs, A. Lagendijk, G. Rempe, and W.L. Vos for support and discussions. This work is partly funded by the Stichting voor Fundamenteel Onderzoek der Materie and STW, which are financially supported by the Nederlandse Organisatie voor Wetenschappelijk Onderzoek (NWO). A.P.M. acknowledges financial support from the European Research Council (grant number 279248). P.W.H.P acknowledges a NWO VICI grant.